\documentclass[aps,prl,twocolumn,showpacs,groupedaddress]{revtex4-1}
\usepackage{graphics}	% for figures (1 of 2)
\usepackage{graphicx}	% for figures (2 of 2)
\usepackage{natbib}		% for combined bibliography entries
\usepackage{dcolumn}	% for some tables
\usepackage{amsmath}	% for math (1 of 2)
\usepackage{amssymb}	% for math (2 of 2)

\begin{document}

\title{Reduced Limit on the Permanent Electric Dipole Moment of $^{199}$Hg}
\author{B. Graner}
\email[]{bgraner@uw.edu}
\author{Y. Chen}
\author{E. G. Lindahl}
\author{B. R. Heckel}
\affiliation{Department of Physics, University of Washington, Seattle, Washington 98195, USA}

\date{\today}
\begin{abstract}
This paper describes the results of the most recent measurement of the permanent electric dipole moment (EDM) of neutral $^{199}$Hg atoms. Fused silica vapor cells containing enriched $^{199}$Hg are arranged in a stack in a common magnetic field. Optical pumping is used to spin-polarize the atoms orthogonal to the applied magnetic field, and the Faraday rotation of near-resonant light is observed to determine an electric-field-induced perturbation to the Larmor precession frequency. Our results for this frequency shift are consistent with zero; we find the corresponding $^{199}$Hg EDM $d_{Hg} = (2.20 \pm 2.75_{stat} \pm 1.48_{syst}) \times 10^{-30} e\cdot \text{cm}$. We use this result to place a new upper limit on the $^{199}$Hg EDM $|d_{Hg}| < 7.4\times 10^{-30} e\cdot \text{cm}$ (95\% C.L.), improving our previous limit by a factor of 4. We also discuss the implications of this result for various $CP$-violating observables as they relate to theories of physics beyond the standard model.
\end{abstract}

\maketitle

The existence of a nonzero permanent electric dipole moment (EDM) oriented along the spin axis of an atom or subatomic particle requires time-reversal symmetry ($T$) violation \cite{1950_Purcell_Ramsey}. By the $CPT$ theorem, $T$-violation implies that $CP$ symmetry must be violated as well. The Standard Model (SM) of particle physics provides two sources of $CP$-violation: a single phase in the CKM matrix \cite{1973_Kobayashi_Maskawa} and $\bar{\theta}_{QCD}$, the coefficient of an allowed $CP$-violating term in the QCD Lagrangian \cite{1976_'t_Hooft_U1_problem_solution_with_Theta_QCD}. However, the CKM phase contribution to any atomic or particle EDM is far below existing experimental sensitivities \cite{Khriplovich_Lamoreaux}, and the measured value of $\bar{\theta}_{QCD}$ is consistent with zero, an apparent anomaly that forms the basis of the Strong $CP$ problem. An atomic EDM may thus provide the first evidence of $CP$-violation in the strong sector, or evidence of $CP$-violating physics beyond the SM \cite{2013_Engel_et_al_EDM_review}. Discovery of any new source of $CP$-violation may also fulfill one of the Sakharov conditions \cite{1967_Sakharov} necessary for a theory of baryogenesis that can reproduce the observed matter excess in the universe \cite{2010_EDM_Baryogenesis}.

There are many ongoing experiments currently searching for a nonzero atomic, electron, or neutron EDM \cite{2015_Ra_EDM, 2014_ACME_eEDM, 2011_YbF_EDM, 2015_ILL_nEDM_gravity_correction}. This paper presents the results of an improved EDM search in the $^{199}$Hg atom \cite{2009_Hg_EDM}. The experiment consists of four (25 mm inner diameter, 10.1 mm tall) vapor cells fabricated from Heraeus Suprasil fused silica and filled with 0.56 atm of CO buffer gas and $\sim$0.5 mg of isotopically-enriched (92\%) $^{199}$Hg, arranged in a stack inside a common magnetic field $\mathbf{B_0}$. The atoms are transverse polarized via optical pumping, and precess with angular frequency $\omega_0 = \gamma\mathbf{B_0}$, where $\gamma=4844 \text{ s}^{-1}/\text{G}$ is the gyromagnetic ratio of $^{199}$Hg. A nonzero EDM, $\mathbf{d} = d_{Hg}\mathbf{I},$ adds a second term to the Hamiltonian $H = -\boldsymbol{\mu}\cdot\mathbf{B} - \mathbf{d}\cdot\mathbf{E}.$ Because the only vector characterizing the system is the nuclear spin ($I=1/2$), any EDM must lie along the spin axis. Degeneracy arguments imply that the EDM can have only one projection onto the spin vector for a given particle or atomic species \cite{Khriplovich_Lamoreaux}. If a two-level atom with a nonzero EDM is placed in parallel fields $\mathbf{B}$, $\mathbf{E}$ and another in antiparallel fields $\mathbf{B}$, $\mathbf{-E}$, the difference in the precession frequency is given by $\hbar\Delta\omega = 4(d_{Hg}E)$.

\begin{figure}
\includegraphics[scale=0.8]{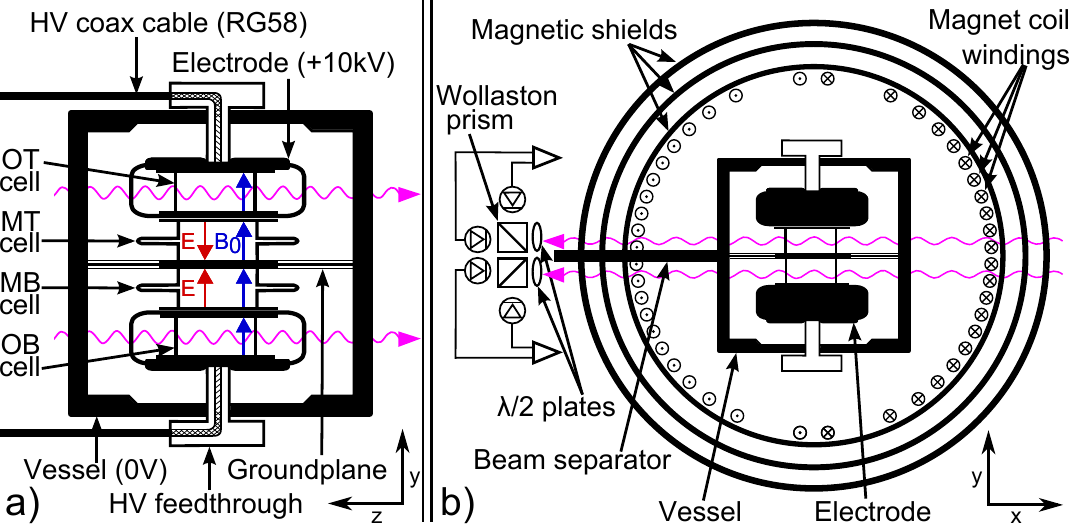}
\caption{Cross-sectional diagrams of the apparatus used to measure the EDM of $^{199}$Hg (not to scale). a) Section of the vessel through the y-z plane showing the high voltage (HV) cables, groundplane, and a cut-away view of the HV electrodes and feedthroughs. b) Section through the x-y plane showing the cylindrical 3-layer magnetic shielding, the cos$(\theta)$ magnet coil windings, and a diagram with 2 of the polarimeters used to observe signals from each of the 4 cells. The laser beams through the outer cells traverse the apparatus along the shield axis (z-axis), while the middle cell beams travel along the x-axis.}
\label{Apparatus_diagram}
\end{figure}

A schematic diagram of the experimental apparatus is given in Fig. \ref{Apparatus_diagram}. The Hg vapor cells are stacked along the axis of the static magnetic field $\mathbf{B_0}$. All four cells are inside a grounded box (called the \textit{vessel}) constructed from anti-static UHMW polyethylene, with a tin(IV) oxide-coated groundplane constructed from 3 layers of 1/16" fused silica dividing the two halves. The two outer cells are seated inside conducting plastic electrodes (maintained at the same potential), so only the inner cells have nonzero electric fields inside (pointing in opposite directions). The outer cells (with $\mathbf{E}=0$) have zero EDM sensitivity and are used as magnetometers and to control sources of systematic error.

\begin{figure}
\includegraphics[scale=0.46]{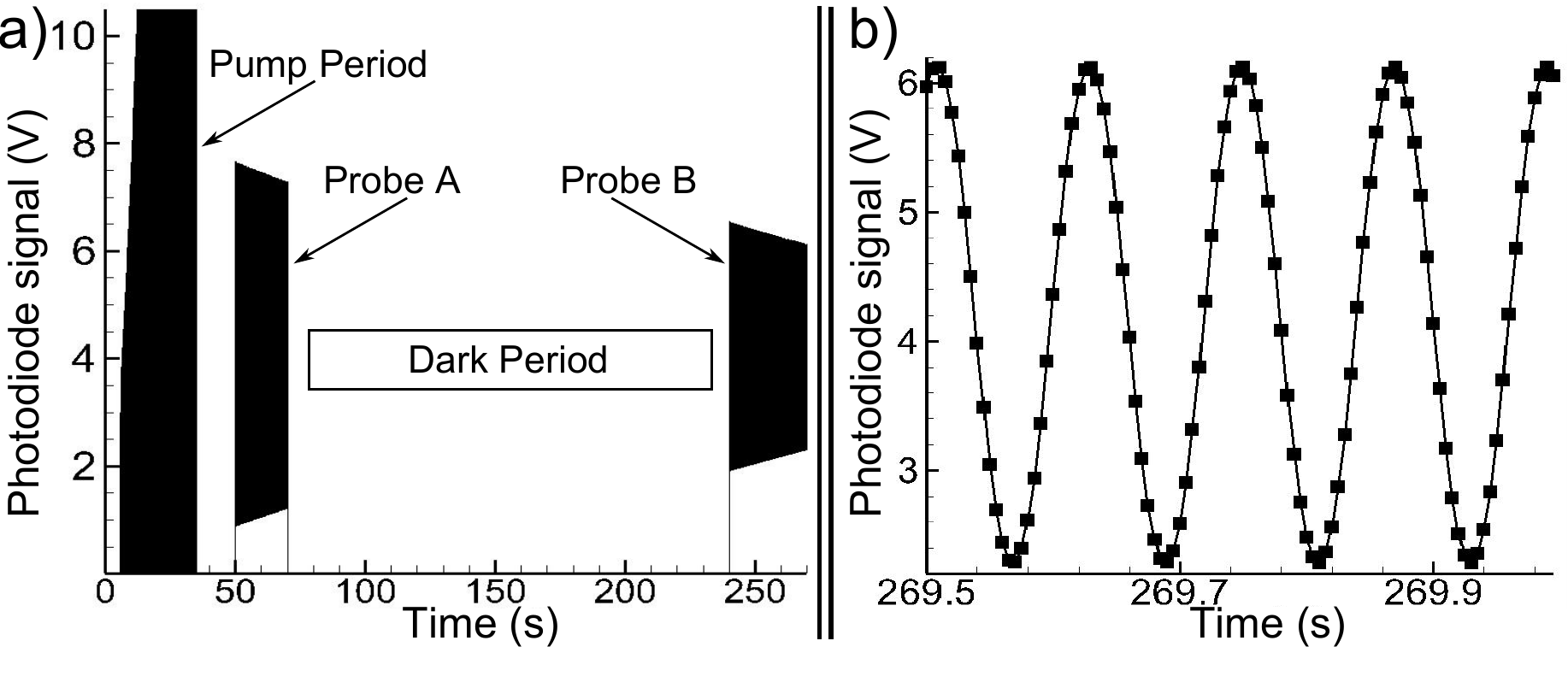}
\caption{The signal obtained from a single photodiode for one pump-probe cycle. a) A complete view of the signal. During optical pumping, the transmission through the cell increases, quickly saturating the detector. The laser power is reduced during the probe periods $A$ and $B$, which are analyzed to extract the phase difference accumulated between cells during the dark period. Individual Larmor oscillations are too rapid to be visible at this scale, but the exponential decay of the signal envelope can be seen. b) An expanded view of the final 500 ms of the data train. The raw data points are connected by straight line segments to guide the eye; no fit is shown.}
\label{Data_timimg}
\end{figure}

A typical pump-probe cycle (illustrated in Fig. \ref{Data_timimg}) lasts 4.5 minutes and consists of 5 parts: A 30 s pump period, a 20 s equilibration period, an initial 20 s probe (period $A$), a 170 s free-precession (the \textit{Dark} period), and a final 30 s probe (period $B$). The dark period is constrained by the $^{199}$Hg spin relaxation time, which varies between 250-600 s between cells. To avoid noise, systematic errors, and cell depolarization caused by the probe light, we sample the precession signals over two brief probe windows. Because phase can be determined more precisely than frequency for a short data train, we extract the EDM precession frequency shift from the phase difference of the middle cells at the end of probe period $A$ and the beginning of probe period $B$.

During the pump phase, the laser light is tuned to the center of the $F=1/2 \rightarrow F=1/2$ component of the $^1S_0\rightarrow{}^{3}P_1$ transition and circularly polarized using a $\lambda/4$ waveplate. To coherently polarize atoms transverse to $\mathbf{B_0}$, the pump light is chopped in sync with the Larmor frequency $\omega_0$ by a rotating chopper wheel with 40\% duty cycle. Following the pump, the $\lambda/4$ waveplate and chopper wheel are removed from the beam path using pneumatic arms. The atoms are probed with linearly polarized laser light which is detuned $\sim$10 GHz to the blue (halfway between the $F=1/2$ and $F=3/2$ components of the excited state) and attenuated to 0.1 times the pump intensity. 

Precession signals for each cell are derived from the Faraday rotation angle of the input probe light polarization, $\theta(t)$, which is proportional to the dot product of the atomic magnetization vector and the light propagation vector: $\theta(t) = \theta_0 \sin(\omega t + \phi) e^{-t/\tau}$ where $\tau$ is the magnetization relaxation time. The output beam from each cell is sent to a balanced polarimeter, where a $\lambda/2$ waveplate rotates the average polarization vector of the light and a Wollaston prism separates it into $s$ and $p$ components. Each component is detected by a UV-enhanced Si photodiode, digitized at a sampling frequency of 2 kHz, and written to disk at 200 Hz after averaging 10 samples. With the $\lambda/2$ waveplate properly set, the time-averaged $s$ and $p$ intensities for each cell are equal, and the cell precession signal becomes
\begin{equation}
S(t) = \frac{I_s(t)-I_p(t)}{I_s(t)+I_p(t)} = \sin{2\theta} \approx 2\theta_0 \sin(\omega t + \phi) e^{-t/\tau}.
\end{equation} 

For each pair of cells $m,n$, we measure the phase difference $\Delta \phi_{mn}$ at the end (beginning) of probe period $A$ ($B$) using phase-sensitive detection. The magnitude of $\bf B_0$ is tuned to give an average precession frequency $\omega_0 = 2\pi \times 8.33 $ s$^{-1}$, and the sampling frequency is 200 Hz, so the digitized signals have 24 points per Larmor cycle: $t_{i+1}-t_i = 5$ ms $=\pi/(12\omega)$. Then $S(t_{i\pm 6}) = \pm 2\theta_0 \cos(\omega t + \phi) e^{-t_{i \pm 6}/\tau}$ and $N^2(t_i) = S^2(t_i) -S(t_{i+6}) S(t_{i-6}) = 4\theta_0^2 e^{-2t_i/\tau}$. With $S'(t_i) = S(t_{i+6}) -S(t_{i-6})$, our beat signal is:
\begin{equation}
2\sin(\Delta \omega_{mn} t_i +\Delta \phi_{mn}) = \frac{S_m(t_i)S'_n(t_i)-S_n(t_i)S'_m(t_i)}{N_m(t_i)N_n(t_i)}.
\end{equation}
Defining $t_i = 0$ at the end of period $A$ or the beginning of period $B$, a least squares fit to $\Delta \omega_{mn} t_i +\Delta \phi_{mn}$ gives us $\Delta \phi_{mn}^{A,B}$. The average dark frequency difference between two cells is $\Delta \omega_{mn}^D = (\Delta \phi_{mn}^B -\Delta \phi_{mn}^A)/\Delta t^D,$ where $\Delta t^D$ is typically 170 s. Because the pump beam strongly alters the spatial distribution of polarization, beat signals obtained immediately after the pump period exhibit substantial nonlinear behavior. The beam is blocked for 20 s between the end of the pump and the beginning of probe period $A$ to ensure the atoms within the volume of the laser beam are in equilibrium with the average phase of atoms throughout the cell.

For any pair of cells, the signature of an EDM is the correlation between $\Delta \omega_{mn}^D$ and the difference in the electric field. During EDM data runs, the high voltage (HV) polarity is reversed between each pump-probe cycle, so in the absence of any noise sources the middle cell frequency difference $\Delta \omega^D_{MT-MB}$ would have an opposite sign for each successive measurement. In reality, $\Delta \omega^D_{MT-MB}$ is also sensitive to fluctuations in the gradients of $\mathbf{B_0}$. To reduce the impact of ambient magnetic field gradient noise, we use the outer cells as magnetometers and define our EDM signal as $\Delta \omega_{EDM} = \Delta \omega^D_{MT-MB}- k\Delta \omega^D_{OT-OB},$ where $k$ is the coefficient that minimizes the variance of the HV-correlated part of $\Delta \omega_{EDM}$ within each daily set of measurements. Runs with small values for $k$ reflect a low level of gradient noise (which is common to $\Delta \omega^D_{MT-MB}$ and $\Delta \omega^D_{OT-OB}$) relative to the statistical uncertainty in $\Delta \omega^D_{OT-OB}$, which simply adds noise to $\Delta \omega_{EDM}$ if $k > 0$ and gradient noise is absent. When gradient noise dominates, the value of $k$ goes to $1/3$ for maximum common-mode noise rejection. Throughout the data set, the value of $k$ varied from 0.18 to 0.33 with an average of 0.25.

\begin{figure}
\includegraphics[scale=0.36]{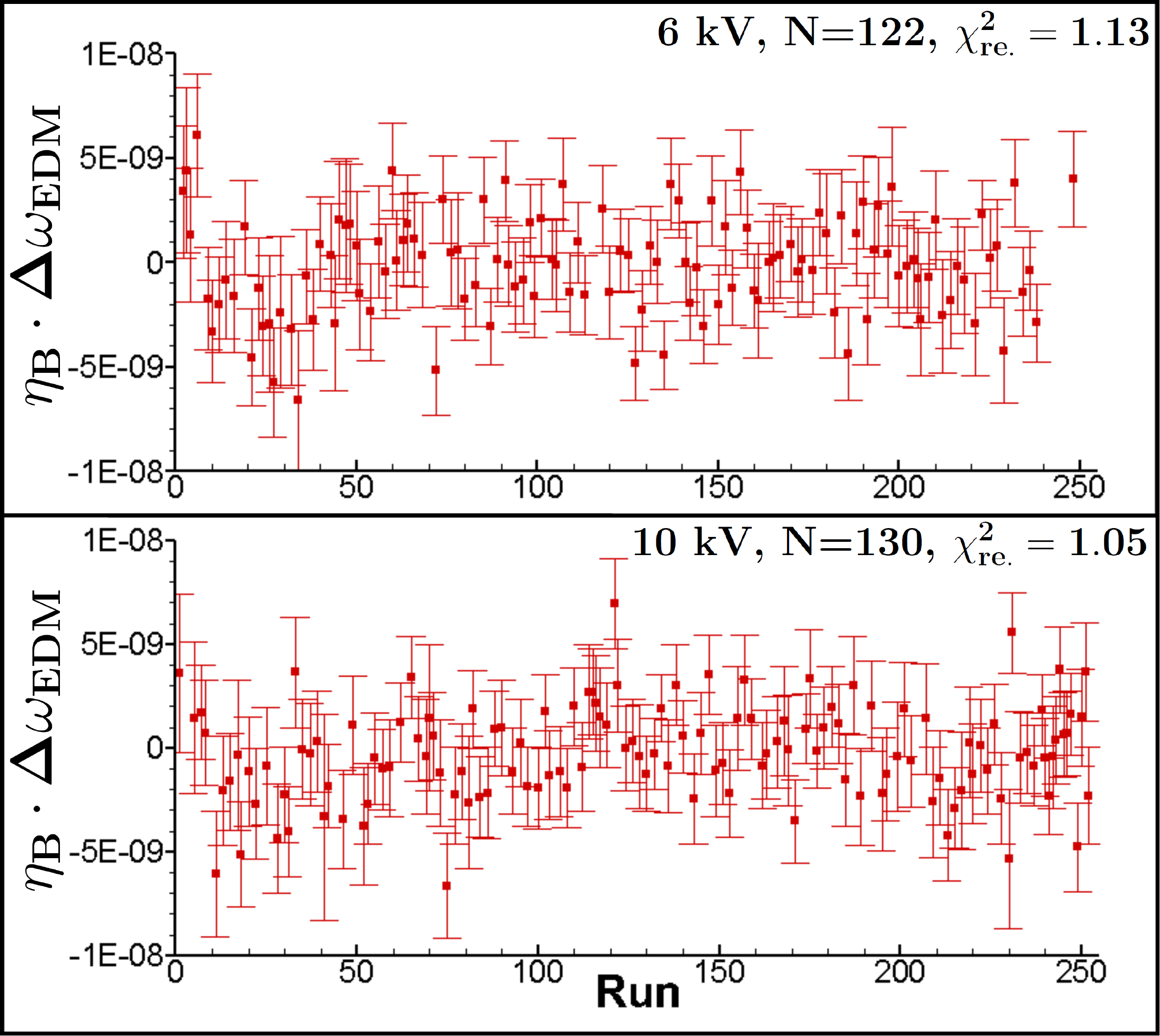}
\caption{The measured EDM frequency shift $\eta_{\mathbf{B}}\cdot\Delta \omega_{EDM}$ for each of the 252 runs in the final (reduced) data set. Each run is plotted on either the top or bottom chart. Top: the frequency shifts measured between $\pm 6$ kV/cm. Bottom: shifts measured between $\pm 10$ kV/cm.}
\label{Combo_Data}
\end{figure}

During normal data taking, the pattern of HV reversals is alternated between $\pm 10$ kV or $\pm 6$ kV each day to check the scaling of an EDM signal with the strength of $\mathbf{E}$. The magnet coil current is reversed every other day to reduce the effect of systematics which depend on the HV but do not change sign with $\mathbf{B_0}$ (as a real EDM signal would). To average data across multiple daily runs, we define $\eta_{\mathbf{B}} = \mathbf{B}_0 \cdot \hat{y}/|\mathbf{B}_0|$ and take $\eta_{\mathbf{B}}\cdot\Delta \omega_{EDM}$ as our EDM-sensitive frequency channel. The results for $\eta_{\mathbf{B}}\cdot\Delta \omega_{EDM}$ from each run are plotted in Fig. \ref{Combo_Data}.

A set of 16-24 data runs with several cycles through the four $\mathbf{B_0}|\mathbf{E}|$ values defines a \textit{sequence}. The full EDM data set is comprised of 284 daily runs divided into 12 sequences. Between sequences, the vessel is opened and the vapor cells are permuted through the various positions. The 3 vapor cells with the longest spin relaxation times are used in the EDM-sensitive middle cell positions. Each of these cells occupied the middle top (MT) and middle bottom (MB) positions for 4 sequences each. 

To make data cuts without biasing the result, the frequency difference of each cell pair is added to a computer-generated blind offset, proportional to the $\mathbf{E}$-field difference between the cells. The sign of the blind offset changes with the direction of $\mathbf{B_0}$, and the value is randomized with each new sequence. After data cuts of individual pump-probe cycles are made within a sequence (based on anomalous behavior on non-EDM data channels), the sequence is reanalyzed with a blind offset common to all runs to enable comparisons across sequences.

During data taking, HV-correlated frequency shifts would sometimes appear on the EDM-insensitive outer cell frequency difference. When this occurred, additional data runs were taken (63 in total) to investigate the source of the anomalous signal. Small, HV-correlated motions of the vapor cells in magnetic field gradients were identified as the cause. To reduce the potential feedthrough of the cell motion onto the EDM signal, 32 runs were cut from the EDM data set based on two criteria: 
\begin{enumerate}
\item $|\Delta\omega_{OT-OB}| > 2.0 \sigma$ or $2.0 \times 10^{-8}$ s$^{-1}$
\item $|\Delta\omega_{(OT-MT)+(OB-MB)}| > 3.0 \sigma$ or $1.5 \times 10^{-8}$ s$^{-1}$ 
\end{enumerate}
The cuts are based only on EDM-insensitive channels (with zero average $\mathbf{E}$ and no blind offset). The first channel is sensitive to cell motion in a linear field gradient $\partial B_y/\partial y.$ The second channel is equivalent to the difference between the average frequencies of the inner and outer cells and can indicate cell motion coupled to the second derivative $\partial^2B_y/\partial y^2.$ After the final cuts, our data set consists of 252 runs encompassing $\sim$65,000 frequency difference measurements to be analyzed for an EDM. The 32 cut runs and 63 auxiliary runs are used to set limits on the systematic error associated with cell motion.

\begin{table}															% h = 'here'
\caption{Measured HV-correlated frequency shifts for various field configurations. Entries for $\Delta\omega$ are specified in units of $10^{-11}$ (kV$\cdot$s/cm)$^{-1}$.}
\begin{tabular}{cccc}													% centered columns (4 columns) 
\hline \hline	 														% inserts double horizontal lines 
Voltage & $\mathbf{B_0} \cdot \hat{y}$ & $\eta_{\mathbf{B}}\cdot\Delta\omega_{EDM}$ & $\eta_{\mathbf{B}}\cdot\Delta\omega_{OT-OB}$ \\ %header
\hline																	% inserts single line
$\pm$10 kV & +10 mG & $(1.16\pm2.6)$ & $(-8.8\pm9.9)$ \\ % body 
$\pm$10 kV & -10 mG & $(-3.15\pm2.8)$ & $(-11.5\pm10.4)$ \\ % body 
$\pm$ 6 kV & +10 mG & $(3.69\pm5.0)$ & $(5.9\pm16.7)$ \\ % body 
$\pm$ 6 kV & -10 mG & $(-8.56\pm4.7)$ & $(-29.1\pm17.1)$ \\ % body 
\hline \hline
\end{tabular} 
\label{Results} 														% used to refer to this table 
\end{table} 

Table \ref{Results} summarizes the results for each HV value and magnetic field direction. We take the average over the two magnetic field directions for both 10 kV and 6 kV and then take the inverse error squared weighted average of the 10 kV and 6 kV results to find our EDM frequency shift $\eta_{\mathbf{B}}\cdot\Delta\omega_{EDM} = (-1.34 \pm 1.67) \times 10^{-11}$ (kV$\cdot$s/cm)$^{-1},$ which gives us an EDM $\mathbf{d}_{Hg} = (2.20 \pm 2.75_{stat}) \times 10^{-30}$ $e\cdot$ cm.

Table \ref{Systematics} gives the systematic error budget of our measurements. The dominant contribution comes from the effect of HV-correlated \textit{Axial Cell Motion.} Detailed maps of the magnetic field gradients at the vapor cells revealed a lab-fixed gradient in $\mathbf{B_0}$ along the axial magnetic shield direction $\hat{z}$. Motion of the vessel or the cells along $\hat{z}$ caused by applying HV would create a non-zero HV-correlated $\Delta \omega$ for any pair of cells. Because the field gradient in $\hat{z}$ does not reverse sign with $\mathbf{B_0}$, the change in a cell's Larmor frequency under cell motion will change sign as $\mathbf{B}_0$ is reversed, leading to a HV-correlated frequency difference between cells with the $\mathbf{E}\cdot\mathbf{B}_0$ symmetry properties of an EDM. By comparing the EDM extracted from the 95 excluded (ex.) runs for which $\Delta \omega_{OT-OB}$ was large to the reduced EDM data set runs, we find:
\begin{equation} 
\dfrac{\eta_{\mathbf{B}}\cdot(\Delta \omega^{ex.}_{EDM}-\Delta \omega_{EDM})}{\eta_{\mathbf{B}}\cdot(\Delta \omega^{ex.}_{OT-OB} - \Delta \omega_{OT-OB})} = (1.6 \pm 5.7)\times 10^{-2}
\end{equation}
which we take as the projection of a $\Delta \omega_{OT-OB}$ cell motion signal onto the EDM channel. This projection times the combined $\eta_{\mathbf{B}}\cdot\Delta \omega_{OT-OB} = (1.04 \pm 0.62)\times10^{-10}$ (kV$\cdot$s/cm)$^{-1}$ gives the top systematic error in Table \ref{Systematics}.

The \textit{Radial Cell Motion} systematic refers to motion in both directions ($\hat{x}$ and $\hat{y}$) orthogonal to the shield axis. The measured field gradients in $\hat{x}$ and $\hat{y}$ were found to reverse to within 4.4\% when the $\mathbf{B_0}$ coil current was reversed. HV-correlated motion along these axes would thus generate a non-zero value for $\Delta \omega_{EDM}$ that, unlike a true EDM signal, would not change sign under $\mathbf{B_0}$ reversal. Table \ref{Results} shows that the non-reversing component of $\Delta \omega_{EDM}$ (half the difference between the two magnetic field directions) is $(3.1 \pm 1.7) \times 10^{-11}$ (kV$\cdot$s/cm)$^{-1}$. A signal of this size could be generated by a HV-correlated motion as small as 2 nm and the projection onto the true EDM channel would be 4.4\% as large.

The \textit{Leakage Current} systematic refers to electrical currents flowing from the HV electrodes to ground along the walls of the vapor cells. A helical current path around a vapor cell would create a magnetic field that adds linearly to $\mathbf{B_0}$, producing a Larmor frequency shift with the same field-dependence as an EDM. Currents from the vessel walls, the HV cables, and various segments of the groundplane were continuously monitored by a set of $0.01$ V/pA transimpedance amplifiers. The measured currents to the top and bottom groundplanes were each $\le 40$ fA, averaged over both probe and dark periods; roughly $30\%$ of these currents were displacement currents from charge accumulation in the HDPE HV feedthroughs. We use 40 fA for an upper limit on the leakage currents and follow the ``worst-case current path'' of 1/2 turn around each of the middle cells to determine the impact on $\Delta\omega_{EDM}$. Dividing by $\sqrt{3}$ to account for the average of 3 independent vapor cells, our total leakage current systematic is 10 times smaller than our previous EDM experiment \cite{2013_Hg_EDM_PRA}. The groundplane in \cite{2013_Hg_EDM_PRA} was coated with Au, with a work function (5.1-5.5 eV) close to the 254 nm photon energy, creating photoelectrons from scattered light. The new groundplane is coated with SnO$_2$.

\begin{table}	 														% h = 'here'
\caption{Systematic effects on the measured EDM value. All entries are specified in units of ($10^{-31} e\cdot \text{cm}$). The final value for the systematic error bar is the quadrature sum of the listed effects, $1.48\times10^{-30} e\cdot \text{cm}$.}
\begin{tabular}{cc|cc}													% centered columns (4 columns) 
\hline \hline															% inserts double horizontal lines 
Effect & Syst. & Effect & Syst. \\ 										% table header
\hline																	% inserts single line
Axial Cell Motion 		&	12.6	&	Parameter Correlations							& 2.33	\\ 	% body
Radial Cell Motion 		&	3.36	&	$\mathbf{v}\times\mathbf{E}/c$ Fields			& 2.29	\\ 	% body 
Leakage Currents		&	5.02	&	Charging Currents								& 1.83	\\ 	% body 
$\mathbf{E}^2$ Effects	&	3.04	&	Geometric Phase									& 0.06	\\ 	% body
\hline \hline
\end{tabular} 
\label{Systematics} 													% used to refer to this table 
\end{table} 

The \textit{$\mathbf{E}^2$ Effects} systematic refers to any mechanism that may couple a small difference in the magnitude of $\mathbf{E}$ between the two polarities to the measured frequency difference. After each EDM data run (with a $+-+-$ HV sequence), shorter data runs (typically 30 pump-probe cycles) were taken with a $+0-0$ HV sequence to test for frequency shifts that scale as $|\mathbf{E}|$. For each sequence, the scalar Stark shift was measured to obtain the difference in magnitude of $\mathbf{E}$ for the two polarities of the HV: $r=(\mathbf{E}^2_+-\mathbf{E}^2_-)/(\mathbf{E}^2_++\mathbf{E}^2_-)\leq 0.02$. The product of $r$ and the frequency difference between scans with the HV on and off, $\eta_{\mathbf{B}}\cdot\Delta\omega_{EDM}(|10 \text{kV}|-|0 \text{kV}|),$ is averaged over all sequences to get the systematic error in Table \ref{Systematics}. The remaining entries in Table \ref{Systematics} are derived following the same methods used in \cite{2013_Hg_EDM_PRA}. Combining the statistical and systematic errors gives our final result
\begin{equation}
d_{Hg} = (2.20 \pm 2.75_{stat} \pm 1.48_{syst}) \times 10^{-30} e\cdot \text{cm,}
\label{EDM_value}
\end{equation}
from which we set a 95\% confidence limit
\begin{equation}
|d_{Hg}| < 7.4\times 10^{-30} e\cdot \text{cm}.
\label{EDM_limit}
\end{equation}

Theoretical interpretations of this limit begin with consideration of the Schiff moment $\mathbf{S}_{Hg},$ the leading-order $P,T$-violating nuclear moment not completely screened by the electron cloud \cite{1963_Schiff, Khriplovich_Lamoreaux, 2004_Ginges_Flambaum_Fund._Symmetries_in_Atoms, 2013_Engel_et_al_EDM_review}. To obtain a limit on $\mathbf{S}_{Hg}$, we average the calculated EDM contributions from \cite{2015_Singh_and_Sahoo_Hg_Schiff_Moment, 2014_Radziute_et_al_DHF_EDM, 2009_Latha_et_al_Hg_EDM, *2015_Latha_et_al_Hg_EDM_Correction, 2009_Dzuba_Flambaum_Porsev_EDMvSM, 2002_Dzuba_et._al._EDMvsSM}: $\mathbf{d}_{Hg} = -2.4\times 10^{-4} \mathbf{S}_{Hg}/\text{fm}^{2}.$ Using Eq. \ref{EDM_limit}, we can set a limit 
\begin{equation}
|S_{Hg}| < 3.1\times 10^{-13} e\cdot \text{fm}^3 \text{(95\% C.L.).}
\end{equation}
We use this value to set limits on other $CP$-violating quantities of interest in Table \ref{Limits}. Limits on the nucleon EDMs $\mathbf{d}_{n,p}$ are derived from the associated contributions to $\mathbf{S}_{Hg}$ in an RPA calculation with core-polarization \cite{2003_Dmitriev_and_Sen'kov_Hg_Schiff_Moment}, which yielded $\mathbf{S}_{Hg} = (1.9 \mathbf{d}_n + 0.2 \mathbf{d}_p) \text{fm}^2$, with a 30\% uncertainty in the $\mathbf{d}_p$ contribution which is reflected in our limit. The $\pi NN$ coupling constants $\bar{g}_{0,1,2}$ parameterize the isoscalar, isovector, and isotensor components of the $CP$-violating nucleon-nucleon interaction, respectively. However, there is considerable disagreement between various calculations of $\mathbf{S}_{Hg}(\bar{g}_0,\bar{g}_1,\bar{g}_2)$. To set limits on $\bar{g}_{0,1,2}$, we use the quoted best values for $^{199}$Hg from the recent review \cite{2013_Engel_et_al_EDM_review}. Note that the calculation has a sign ambiguity for the value of $\bar{g}_1$. We also set a limit on $\bar{\theta}_{QCD}$ from the relation $|\bar{g}_0| = 15.5\times 10^{-3}|\bar{\theta}_{QCD}|$ \cite{2015_deVries_Mereghetti_Walker_CP_violation_in_CPT, 2015_Bsaisou_et_al_Nucleon_EDMs_in_chiral_EFT}, although comprehensive lattice calculations of $\mathbf{d}_n(\bar{\theta}_{QCD})$ may provide a tighter bound \cite{2015_Guo_et_al_nEDM_from_lattice_QCD}. A limit on the combined chromo-EDM of the up and down quarks is determined by $\bar{g}_1=2\times10^{-14}\text{cm}^{-1}(\widetilde{d}_u-\widetilde{d}_d)$ \cite{2002_Pospelov_quark_chromo_EDM}.

\begin{table}[h]														% h = 'here'
\caption{Limits on $CP$-violating observables from the $^{199}$Hg EDM limit. Each limit is based on the assumption that it is the sole contribution to the atomic EDM. In principle, the result for $\mathbf{d}_n$ supercedes \cite{2015_ILL_nEDM_gravity_correction} as the best neutron EDM limit.}
\begin{tabular}{cccc}													% centered columns (3 columns) 
\hline \hline	 														% inserts double horizontal lines 
Quantity & Expression & Limit & Ref. \\ 								% table header
\hline 																	% horizontal line underneath header
\\[-2.0ex]			% inserts extra line and subtracts space (w/o extra space, hline cuts through superscripts on the following line)
$\mathbf{d}_n$	& $\mathbf{S}_{Hg}/(1.9 \text{ fm}^2)$ 		& $1.6\times 10^{-26}$ $e\cdot\text{cm}$ & \cite{2003_Dmitriev_and_Sen'kov_Hg_Schiff_Moment}\\
$\mathbf{d}_p$	& $1.3\times\mathbf{S}_{Hg}/(0.2 \text{ fm}^2)$	& $2.0\times 10^{-25}$ $e\cdot\text{cm}$ & \cite{2003_Dmitriev_and_Sen'kov_Hg_Schiff_Moment}\\
$\bar{g}_0$ 	& $\mathbf{S}_{Hg}/(0.135$ $e\cdot\text{fm}^3)$ & $2.3\times 10^{-12}$ & \cite{2013_Engel_et_al_EDM_review}\\
$\bar{g}_1$ 	& $\mathbf{S}_{Hg}/(0.27$ $e\cdot\text{fm}^3)$ 	& $1.1\times 10^{-12}$ & \cite{2013_Engel_et_al_EDM_review}\\
$\bar{g}_2$ 	& $\mathbf{S}_{Hg}/(0.27$ $e\cdot\text{fm}^3)$ 	& $1.1\times 10^{-12}$ & \cite{2013_Engel_et_al_EDM_review}\\
$\bar{\theta}_{QCD}$ 	& $\bar{g}_0/0.0155$ 					& $1.5\times 10^{-10}$ & \cite{2015_deVries_Mereghetti_Walker_CP_violation_in_CPT, 2015_Bsaisou_et_al_Nucleon_EDMs_in_chiral_EFT}\\
$(\widetilde{d}_u-\widetilde{d}_d)$ & $\bar{g}_1/(2\times10^{14} \text{cm}^{-1})$ & $5.7\times 10^{-27} \text{ cm}$& \cite{2002_Pospelov_quark_chromo_EDM}\\
$C_S$ 			& $\mathbf{d}_{Hg}/(5.9 \times10^{-22}$ $e\cdot\text{cm})$ & $1.3\times 10^{-8}$ & \cite{2004_Ginges_Flambaum_Fund._Symmetries_in_Atoms}\\
$C_P$			& $\mathbf{d}_{Hg}/(6.0\times10^{-23}$ $e\cdot\text{cm)}$ & $1.2\times 10^{-7}$ & \cite{2004_Ginges_Flambaum_Fund._Symmetries_in_Atoms}\\
$C_T$			& $\mathbf{d}_{Hg}/(4.89 \times10^{-20}$ $e\cdot\text{cm})$ & $1.5\times 10^{-10}$ & see text\\						
\hline \hline															% inserts double horizontal lines
\end{tabular} 
\label{Limits} 															% used to refer to this table 
\end{table}

Our result can also be used to place limits on $P,T$-odd scalar, pseudoscalar, and tensor electron-nucleon interactions (described by $C_S$, $C_P$, and $C_T$) which may induce an atomic EDM independent of the Schiff moment. In $^{199}$Hg the tensor interaction is expected to dominate. Many recent calculations of the tensor coefficient $C_T$ have been performed; the average result of \cite{2015_Singh_and_Sahoo_Hg_Schiff_Moment, 2014_Radziute_et_al_DHF_EDM, 2009_Dzuba_Flambaum_Porsev_EDMvSM, 2009_Latha_et_al_Hg_EDM, 2008_Latha_et_al_Core_Polrization_EDM} is $\mathbf{d}_{Hg} = -4.89\times 10^{-20} C_T \langle\sigma_N\rangle e\cdot \text{cm}$. Finally, it should be noted that because there are many potential contributions to an atomic EDM, multiple non-null results in different systems will be necessary to extract unambiguous values for fundamental physics parameters \cite{2015_Chupp_Xe_EDM_Motivation}.

\begin{acknowledgments}
We gratefully acknowledge the contributions of Kyle Matsuda for performing our magnetic field maps and Professor E. N. Fortson for valuable advice. This work was supported by NSF Grant 1306743 and the U.S. Department of Energy Office of Science, Office of Nuclear Physics under Award No. DE-FG02-97ER41020.
\end{acknowledgments}
\bibliography{Hg_EDM_2015_V_4}

\begin{thebibliography}{28}%
\makeatletter
\providecommand \@ifxundefined [1]{%
 \@ifx{#1\undefined}
}%
\providecommand \@ifnum [1]{%
 \ifnum #1\expandafter \@firstoftwo
 \else \expandafter \@secondoftwo
 \fi
}%
\providecommand \@ifx [1]{%
 \ifx #1\expandafter \@firstoftwo
 \else \expandafter \@secondoftwo
 \fi
}%
\providecommand \natexlab [1]{#1}%
\providecommand \enquote  [1]{``#1''}%
\providecommand \bibnamefont  [1]{#1}%
\providecommand \bibfnamefont [1]{#1}%
\providecommand \citenamefont [1]{#1}%
\providecommand \href@noop [0]{\@secondoftwo}%
\providecommand \href [0]{\begingroup \@sanitize@url \@href}%
\providecommand \@href[1]{\@@startlink{#1}\@@href}%
\providecommand \@@href[1]{\endgroup#1\@@endlink}%
\providecommand \@sanitize@url [0]{\catcode `\\12\catcode `\$12\catcode
  `\&12\catcode `\#12\catcode `\^12\catcode `\_12\catcode `\%12\relax}%
\providecommand \@@startlink[1]{}%
\providecommand \@@endlink[0]{}%
\providecommand \url  [0]{\begingroup\@sanitize@url \@url }%
\providecommand \@url [1]{\endgroup\@href {#1}{\urlprefix }}%
\providecommand \urlprefix  [0]{URL }%
\providecommand \Eprint [0]{\href }%
\providecommand \doibase [0]{http://dx.doi.org/}%
\providecommand \selectlanguage [0]{\@gobble}%
\providecommand \bibinfo  [0]{\@secondoftwo}%
\providecommand \bibfield  [0]{\@secondoftwo}%
\providecommand \translation [1]{[#1]}%
\providecommand \BibitemOpen [0]{}%
\providecommand \bibitemStop [0]{}%
\providecommand \bibitemNoStop [0]{.\EOS\space}%
\providecommand \EOS [0]{\spacefactor3000\relax}%
\providecommand \BibitemShut  [1]{\csname bibitem#1\endcsname}%
\let\auto@bib@innerbib\@empty
%</preamble>
\bibitem [{\citenamefont {Purcell}\ and\ \citenamefont
  {Ramsey}(1950)}]{1950_Purcell_Ramsey}%
  \BibitemOpen
  \bibfield  {author} {\bibinfo {author} {\bibfnamefont {E.~M.}\ \bibnamefont
  {Purcell}}\ and\ \bibinfo {author} {\bibfnamefont {N.~F.}\ \bibnamefont
  {Ramsey}},\ }\href@noop {} {\bibfield  {journal} {\bibinfo  {journal} {Phys.
  Rev.}\ }\textbf {\bibinfo {volume} {78}},\ \bibinfo {pages} {807} (\bibinfo
  {year} {1950})}\BibitemShut {NoStop}%
\bibitem [{\citenamefont {Kobayashi}\ and\ \citenamefont
  {Maskawa}(1973)}]{1973_Kobayashi_Maskawa}%
  \BibitemOpen
  \bibfield  {author} {\bibinfo {author} {\bibfnamefont {M.}~\bibnamefont
  {Kobayashi}}\ and\ \bibinfo {author} {\bibfnamefont {T.}~\bibnamefont
  {Maskawa}},\ }\href@noop {} {\bibfield  {journal} {\bibinfo  {journal} {Prog.
  Theor. Phys.}\ }\textbf {\bibinfo {volume} {49}},\ \bibinfo {pages} {652}
  (\bibinfo {year} {1973})}\BibitemShut {NoStop}%
\bibitem [{\citenamefont
  {'t~Hooft}(1976)}]{1976_'t_Hooft_U1_problem_solution_with_Theta_QCD}%
  \BibitemOpen
  \bibfield  {author} {\bibinfo {author} {\bibfnamefont {G.}~\bibnamefont
  {'t~Hooft}},\ }\href@noop {} {\bibfield  {journal} {\bibinfo  {journal}
  {Phys. Rev. Lett.}\ }\textbf {\bibinfo {volume} {37}},\ \bibinfo {pages} {8}
  (\bibinfo {year} {1976})}\BibitemShut {NoStop}%
\bibitem [{\citenamefont {Khriplovich}\ and\ \citenamefont
  {Lamoreaux}(1997)}]{Khriplovich_Lamoreaux}%
  \BibitemOpen
  \bibfield  {author} {\bibinfo {author} {\bibfnamefont {I.~B.}\ \bibnamefont
  {Khriplovich}}\ and\ \bibinfo {author} {\bibfnamefont {S.~K.}\ \bibnamefont
  {Lamoreaux}},\ }\href@noop {} {\emph {\bibinfo {title} {CP Violation Without
  Strangeness}}}\ (\bibinfo  {publisher} {Springer-Verlag},\ \bibinfo {year}
  {1997})\BibitemShut {NoStop}%
\bibitem [{\citenamefont {Engel}\ \emph {et~al.}(2013)\citenamefont {Engel},
  \citenamefont {Ramsey-Musolf},\ and\ \citenamefont {van
  Kolck}}]{2013_Engel_et_al_EDM_review}%
  \BibitemOpen
  \bibfield  {author} {\bibinfo {author} {\bibfnamefont {J.}~\bibnamefont
  {Engel}}, \bibinfo {author} {\bibfnamefont {M.~J.}\ \bibnamefont
  {Ramsey-Musolf}}, \ and\ \bibinfo {author} {\bibfnamefont {U.}~\bibnamefont
  {van Kolck}},\ }\href@noop {} {\bibfield  {journal} {\bibinfo  {journal}
  {Prog. Part. Nucl. Phys.}\ }\textbf {\bibinfo {volume} {71}},\ \bibinfo
  {pages} {21} (\bibinfo {year} {2013})}\BibitemShut {NoStop}%
\bibitem [{\citenamefont {Sakharov}(1967)}]{1967_Sakharov}%
  \BibitemOpen
  \bibfield  {author} {\bibinfo {author} {\bibfnamefont {A.~D.}\ \bibnamefont
  {Sakharov}},\ }\href@noop {} {\bibfield  {journal} {\bibinfo  {journal} {Sov.
  Phys. JETP Lett.}\ }\textbf {\bibinfo {volume} {5}},\ \bibinfo {pages} {24}
  (\bibinfo {year} {1967})}\BibitemShut {NoStop}%
\bibitem [{\citenamefont {Cirigliano}\ \emph {et~al.}(2010)\citenamefont
  {Cirigliano}, \citenamefont {Li}, \citenamefont {Profumo},\ and\
  \citenamefont {Ramsey-Musolf}}]{2010_EDM_Baryogenesis}%
  \BibitemOpen
  \bibfield  {author} {\bibinfo {author} {\bibfnamefont {V.}~\bibnamefont
  {Cirigliano}}, \bibinfo {author} {\bibfnamefont {Y.}~\bibnamefont {Li}},
  \bibinfo {author} {\bibfnamefont {S.}~\bibnamefont {Profumo}}, \ and\
  \bibinfo {author} {\bibfnamefont {M.~J.}\ \bibnamefont {Ramsey-Musolf}},\
  }\href@noop {} {\bibfield  {journal} {\bibinfo  {journal} {J. High Energy
  Phys.}\ }\textbf {\bibinfo {volume} {01}} (\bibinfo {year} {2010})},\
  \bibinfo {note} {002}\BibitemShut {NoStop}%
\bibitem [{\citenamefont {Parker}\ \emph {et~al.}(2015)\citenamefont {Parker},
  \citenamefont {Dietrich}, \citenamefont {Kalita}, \citenamefont {Lemke},
  \citenamefont {Bailey}, \citenamefont {Bishof}, \citenamefont {Greene},
  \citenamefont {Holt}, \citenamefont {Korsch}, \citenamefont {Lu},
  \citenamefont {Mueller}, \citenamefont {O'Connor},\ and\ \citenamefont
  {Singh}}]{2015_Ra_EDM}%
  \BibitemOpen
  \bibfield  {author} {\bibinfo {author} {\bibfnamefont {R.~H.}\ \bibnamefont
  {Parker}}, \bibinfo {author} {\bibfnamefont {M.~R.}\ \bibnamefont
  {Dietrich}}, \bibinfo {author} {\bibfnamefont {M.~R.}\ \bibnamefont
  {Kalita}}, \bibinfo {author} {\bibfnamefont {N.~D.}\ \bibnamefont {Lemke}},
  \bibinfo {author} {\bibfnamefont {K.~G.}\ \bibnamefont {Bailey}}, \bibinfo
  {author} {\bibfnamefont {M.}~\bibnamefont {Bishof}}, \bibinfo {author}
  {\bibfnamefont {J.~P.}\ \bibnamefont {Greene}}, \bibinfo {author}
  {\bibfnamefont {R.~J.}\ \bibnamefont {Holt}}, \bibinfo {author}
  {\bibfnamefont {W.}~\bibnamefont {Korsch}}, \bibinfo {author} {\bibfnamefont
  {Z.-T.}\ \bibnamefont {Lu}}, \bibinfo {author} {\bibfnamefont
  {P.}~\bibnamefont {Mueller}}, \bibinfo {author} {\bibfnamefont {T.~P.}\
  \bibnamefont {O'Connor}}, \ and\ \bibinfo {author} {\bibfnamefont {J.~T.}\
  \bibnamefont {Singh}},\ }\href@noop {} {\bibfield  {journal} {\bibinfo
  {journal} {Phys. Rev. Lett.}\ }\textbf {\bibinfo {volume} {114}},\ \bibinfo
  {pages} {233002} (\bibinfo {year} {2015})}\BibitemShut {NoStop}%
\bibitem [{\citenamefont {Baron}\ \emph {et~al.}(2014)\citenamefont {Baron},
  \citenamefont {Campbell}, \citenamefont {DeMille}, \citenamefont {Doyle},
  \citenamefont {Gabrielse}, \citenamefont {Gurevich}, \citenamefont {Hess},
  \citenamefont {Hutzler}, \citenamefont {Kirilov}, \citenamefont {Kozyryev},
  \citenamefont {O'Leary}, \citenamefont {Panda}, \citenamefont {Parsons},
  \citenamefont {Petrik}, \citenamefont {Spaun}, \citenamefont {Vutha},\ and\
  \citenamefont {West}}]{2014_ACME_eEDM}%
  \BibitemOpen
  \bibfield  {author} {\bibinfo {author} {\bibfnamefont {J.}~\bibnamefont
  {Baron}}, \bibinfo {author} {\bibfnamefont {W.~C.}\ \bibnamefont {Campbell}},
  \bibinfo {author} {\bibfnamefont {D.}~\bibnamefont {DeMille}}, \bibinfo
  {author} {\bibfnamefont {J.~M.}\ \bibnamefont {Doyle}}, \bibinfo {author}
  {\bibfnamefont {G.}~\bibnamefont {Gabrielse}}, \bibinfo {author}
  {\bibfnamefont {Y.~V.}\ \bibnamefont {Gurevich}}, \bibinfo {author}
  {\bibfnamefont {P.~W.}\ \bibnamefont {Hess}}, \bibinfo {author}
  {\bibfnamefont {N.~R.}\ \bibnamefont {Hutzler}}, \bibinfo {author}
  {\bibfnamefont {E.}~\bibnamefont {Kirilov}}, \bibinfo {author} {\bibfnamefont
  {I.}~\bibnamefont {Kozyryev}}, \bibinfo {author} {\bibfnamefont {B.~R.}\
  \bibnamefont {O'Leary}}, \bibinfo {author} {\bibfnamefont {C.~D.}\
  \bibnamefont {Panda}}, \bibinfo {author} {\bibfnamefont {M.~F.}\ \bibnamefont
  {Parsons}}, \bibinfo {author} {\bibfnamefont {E.~S.}\ \bibnamefont {Petrik}},
  \bibinfo {author} {\bibfnamefont {B.}~\bibnamefont {Spaun}}, \bibinfo
  {author} {\bibfnamefont {A.~C.}\ \bibnamefont {Vutha}}, \ and\ \bibinfo
  {author} {\bibfnamefont {A.~D.}\ \bibnamefont {West}} (\bibinfo
  {collaboration} {The ACME Collaboration}),\ }\href@noop {} {\bibfield
  {journal} {\bibinfo  {journal} {Science}\ }\textbf {\bibinfo {volume}
  {343}},\ \bibinfo {pages} {269} (\bibinfo {year} {2014})}\BibitemShut
  {NoStop}%
\bibitem [{\citenamefont {Hudson}\ \emph {et~al.}(2011)\citenamefont {Hudson},
  \citenamefont {Kara}, \citenamefont {Smallman}, \citenamefont {Sauer},
  \citenamefont {Tarbutt},\ and\ \citenamefont {Hinds}}]{2011_YbF_EDM}%
  \BibitemOpen
  \bibfield  {author} {\bibinfo {author} {\bibfnamefont {J.~J.}\ \bibnamefont
  {Hudson}}, \bibinfo {author} {\bibfnamefont {D.~M.}\ \bibnamefont {Kara}},
  \bibinfo {author} {\bibfnamefont {I.~J.}\ \bibnamefont {Smallman}}, \bibinfo
  {author} {\bibfnamefont {B.~E.}\ \bibnamefont {Sauer}}, \bibinfo {author}
  {\bibfnamefont {M.~R.}\ \bibnamefont {Tarbutt}}, \ and\ \bibinfo {author}
  {\bibfnamefont {E.~A.}\ \bibnamefont {Hinds}},\ }\href@noop {} {\bibfield
  {journal} {\bibinfo  {journal} {Nature}\ }\textbf {\bibinfo {volume} {473}},\
  \bibinfo {pages} {493} (\bibinfo {year} {2011})}\BibitemShut {NoStop}%
\bibitem [{\citenamefont {Pendlebury}\ \emph {et~al.}(2015)\citenamefont
  {Pendlebury}, \citenamefont {Afach}, \citenamefont {Ayres}, \citenamefont
  {Baker}, \citenamefont {Ban}, \citenamefont {Bison}, \citenamefont {Bodek},
  \citenamefont {Burghoff}, \citenamefont {Geltenbort}, \citenamefont {Green},
  \citenamefont {Griffith}, \citenamefont {van~der Grinten}, \citenamefont
  {Gruji\ifmmode~\acute{c}\else \'{c}\fi{}}, \citenamefont {Harris},
  \citenamefont {H\'elaine}, \citenamefont {Iaydjiev}, \citenamefont {Ivanov},
  \citenamefont {Kasprzak}, \citenamefont {Kermaidic}, \citenamefont {Kirch},
  \citenamefont {Koch}, \citenamefont {Komposch}, \citenamefont {Kozela},
  \citenamefont {Krempel}, \citenamefont {Lauss}, \citenamefont {Lefort},
  \citenamefont {Lemi\`ere}, \citenamefont {May}, \citenamefont {Musgrave},
  \citenamefont {Naviliat-Cuncic}, \citenamefont {Piegsa}, \citenamefont
  {Pignol}, \citenamefont {Prashanth}, \citenamefont {Qu\'em\'ener},
  \citenamefont {Rawlik}, \citenamefont {Rebreyend}, \citenamefont
  {Richardson}, \citenamefont {Ries}, \citenamefont {Roccia}, \citenamefont
  {Rozpedzik}, \citenamefont {Schnabel}, \citenamefont {Schmidt-Wellenburg},
  \citenamefont {Severijns}, \citenamefont {Shiers}, \citenamefont {Thorne},
  \citenamefont {Weis}, \citenamefont {Winston}, \citenamefont {Wursten},
  \citenamefont {Zejma},\ and\ \citenamefont
  {Zsigmond}}]{2015_ILL_nEDM_gravity_correction}%
  \BibitemOpen
  \bibfield  {author} {\bibinfo {author} {\bibfnamefont {J.~M.}\ \bibnamefont
  {Pendlebury}}, \bibinfo {author} {\bibfnamefont {S.}~\bibnamefont {Afach}},
  \bibinfo {author} {\bibfnamefont {N.~J.}\ \bibnamefont {Ayres}}, \bibinfo
  {author} {\bibfnamefont {C.~A.}\ \bibnamefont {Baker}}, \bibinfo {author}
  {\bibfnamefont {G.}~\bibnamefont {Ban}}, \bibinfo {author} {\bibfnamefont
  {G.}~\bibnamefont {Bison}}, \bibinfo {author} {\bibfnamefont
  {K.}~\bibnamefont {Bodek}}, \bibinfo {author} {\bibfnamefont
  {M.}~\bibnamefont {Burghoff}}, \bibinfo {author} {\bibfnamefont
  {P.}~\bibnamefont {Geltenbort}}, \bibinfo {author} {\bibfnamefont
  {K.}~\bibnamefont {Green}}, \bibinfo {author} {\bibfnamefont {W.~C.}\
  \bibnamefont {Griffith}}, \bibinfo {author} {\bibfnamefont {M.}~\bibnamefont
  {van~der Grinten}}, \bibinfo {author} {\bibfnamefont {Z.~D.}\ \bibnamefont
  {Gruji\ifmmode~\acute{c}\else \'{c}\fi{}}}, \bibinfo {author} {\bibfnamefont
  {P.~G.}\ \bibnamefont {Harris}}, \bibinfo {author} {\bibfnamefont
  {V.}~\bibnamefont {H\'elaine}}, \bibinfo {author} {\bibfnamefont
  {P.}~\bibnamefont {Iaydjiev}}, \bibinfo {author} {\bibfnamefont {S.~N.}\
  \bibnamefont {Ivanov}}, \bibinfo {author} {\bibfnamefont {M.}~\bibnamefont
  {Kasprzak}}, \bibinfo {author} {\bibfnamefont {Y.}~\bibnamefont {Kermaidic}},
  \bibinfo {author} {\bibfnamefont {K.}~\bibnamefont {Kirch}}, \bibinfo
  {author} {\bibfnamefont {H.-C.}\ \bibnamefont {Koch}}, \bibinfo {author}
  {\bibfnamefont {S.}~\bibnamefont {Komposch}}, \bibinfo {author}
  {\bibfnamefont {A.}~\bibnamefont {Kozela}}, \bibinfo {author} {\bibfnamefont
  {J.}~\bibnamefont {Krempel}}, \bibinfo {author} {\bibfnamefont
  {B.}~\bibnamefont {Lauss}}, \bibinfo {author} {\bibfnamefont
  {T.}~\bibnamefont {Lefort}}, \bibinfo {author} {\bibfnamefont
  {Y.}~\bibnamefont {Lemi\`ere}}, \bibinfo {author} {\bibfnamefont {D.~J.~R.}\
  \bibnamefont {May}}, \bibinfo {author} {\bibfnamefont {M.}~\bibnamefont
  {Musgrave}}, \bibinfo {author} {\bibfnamefont {O.}~\bibnamefont
  {Naviliat-Cuncic}}, \bibinfo {author} {\bibfnamefont {F.~M.}\ \bibnamefont
  {Piegsa}}, \bibinfo {author} {\bibfnamefont {G.}~\bibnamefont {Pignol}},
  \bibinfo {author} {\bibfnamefont {P.~N.}\ \bibnamefont {Prashanth}}, \bibinfo
  {author} {\bibfnamefont {G.}~\bibnamefont {Qu\'em\'ener}}, \bibinfo {author}
  {\bibfnamefont {M.}~\bibnamefont {Rawlik}}, \bibinfo {author} {\bibfnamefont
  {D.}~\bibnamefont {Rebreyend}}, \bibinfo {author} {\bibfnamefont {J.~D.}\
  \bibnamefont {Richardson}}, \bibinfo {author} {\bibfnamefont
  {D.}~\bibnamefont {Ries}}, \bibinfo {author} {\bibfnamefont {S.}~\bibnamefont
  {Roccia}}, \bibinfo {author} {\bibfnamefont {D.}~\bibnamefont {Rozpedzik}},
  \bibinfo {author} {\bibfnamefont {A.}~\bibnamefont {Schnabel}}, \bibinfo
  {author} {\bibfnamefont {P.}~\bibnamefont {Schmidt-Wellenburg}}, \bibinfo
  {author} {\bibfnamefont {N.}~\bibnamefont {Severijns}}, \bibinfo {author}
  {\bibfnamefont {D.}~\bibnamefont {Shiers}}, \bibinfo {author} {\bibfnamefont
  {J.~A.}\ \bibnamefont {Thorne}}, \bibinfo {author} {\bibfnamefont
  {A.}~\bibnamefont {Weis}}, \bibinfo {author} {\bibfnamefont {O.~J.}\
  \bibnamefont {Winston}}, \bibinfo {author} {\bibfnamefont {E.}~\bibnamefont
  {Wursten}}, \bibinfo {author} {\bibfnamefont {J.}~\bibnamefont {Zejma}}, \
  and\ \bibinfo {author} {\bibfnamefont {G.}~\bibnamefont {Zsigmond}},\
  }\href@noop {} {\bibfield  {journal} {\bibinfo  {journal} {Phys. Rev. D}\
  }\textbf {\bibinfo {volume} {92}},\ \bibinfo {pages} {092003} (\bibinfo
  {year} {2015})}\BibitemShut {NoStop}%
\bibitem [{\citenamefont {Griffith}\ \emph {et~al.}(2009)\citenamefont
  {Griffith}, \citenamefont {Swallows}, \citenamefont {Loftus}, \citenamefont
  {Romalis}, \citenamefont {Heckel},\ and\ \citenamefont
  {Fortson}}]{2009_Hg_EDM}%
  \BibitemOpen
  \bibfield  {author} {\bibinfo {author} {\bibfnamefont {W.~C.}\ \bibnamefont
  {Griffith}}, \bibinfo {author} {\bibfnamefont {M.~D.}\ \bibnamefont
  {Swallows}}, \bibinfo {author} {\bibfnamefont {T.~H.}\ \bibnamefont
  {Loftus}}, \bibinfo {author} {\bibfnamefont {M.~V.}\ \bibnamefont {Romalis}},
  \bibinfo {author} {\bibfnamefont {B.~R.}\ \bibnamefont {Heckel}}, \ and\
  \bibinfo {author} {\bibfnamefont {E.~N.}\ \bibnamefont {Fortson}},\
  }\href@noop {} {\bibfield  {journal} {\bibinfo  {journal} {Phys. Rev. Lett.}\
  }\textbf {\bibinfo {volume} {102}},\ \bibinfo {pages} {101601} (\bibinfo
  {year} {2009})}\BibitemShut {NoStop}%
\bibitem [{\citenamefont {Swallows}\ \emph {et~al.}(2013)\citenamefont
  {Swallows}, \citenamefont {Loftus}, \citenamefont {Griffith}, \citenamefont
  {Heckel}, \citenamefont {Fortson},\ and\ \citenamefont
  {Romalis}}]{2013_Hg_EDM_PRA}%
  \BibitemOpen
  \bibfield  {author} {\bibinfo {author} {\bibfnamefont {M.~D.}\ \bibnamefont
  {Swallows}}, \bibinfo {author} {\bibfnamefont {T.~H.}\ \bibnamefont
  {Loftus}}, \bibinfo {author} {\bibfnamefont {W.~C.}\ \bibnamefont
  {Griffith}}, \bibinfo {author} {\bibfnamefont {B.~R.}\ \bibnamefont
  {Heckel}}, \bibinfo {author} {\bibfnamefont {E.~N.}\ \bibnamefont {Fortson}},
  \ and\ \bibinfo {author} {\bibfnamefont {M.~V.}\ \bibnamefont {Romalis}},\
  }\href@noop {} {\bibfield  {journal} {\bibinfo  {journal} {Phys. Rev. A}\
  }\textbf {\bibinfo {volume} {87}},\ \bibinfo {pages} {012102} (\bibinfo
  {year} {2013})}\BibitemShut {NoStop}%
\bibitem [{\citenamefont {Schiff}(1963)}]{1963_Schiff}%
  \BibitemOpen
  \bibfield  {author} {\bibinfo {author} {\bibfnamefont {L.~I.}\ \bibnamefont
  {Schiff}},\ }\href@noop {} {\bibfield  {journal} {\bibinfo  {journal} {Phys.
  Rev.}\ }\textbf {\bibinfo {volume} {132}},\ \bibinfo {pages} {2194} (\bibinfo
  {year} {1963})}\BibitemShut {NoStop}%
\bibitem [{\citenamefont {Ginges}\ and\ \citenamefont
  {Flambaum}(2004)}]{2004_Ginges_Flambaum_Fund._Symmetries_in_Atoms}%
  \BibitemOpen
  \bibfield  {author} {\bibinfo {author} {\bibfnamefont {J.~S.~M.}\
  \bibnamefont {Ginges}}\ and\ \bibinfo {author} {\bibfnamefont {V.~V.}\
  \bibnamefont {Flambaum}},\ }\href@noop {} {\bibfield  {journal} {\bibinfo
  {journal} {Phys. Rep.}\ }\textbf {\bibinfo {volume} {397}},\ \bibinfo {pages}
  {63} (\bibinfo {year} {2004})}\BibitemShut {NoStop}%
\bibitem [{\citenamefont {Singh}\ and\ \citenamefont
  {Sahoo}(2015)}]{2015_Singh_and_Sahoo_Hg_Schiff_Moment}%
  \BibitemOpen
  \bibfield  {author} {\bibinfo {author} {\bibfnamefont {Y.}~\bibnamefont
  {Singh}}\ and\ \bibinfo {author} {\bibfnamefont {B.~K.}\ \bibnamefont
  {Sahoo}},\ }\href@noop {} {\bibfield  {journal} {\bibinfo  {journal} {Phys.
  Rev. A}\ }\textbf {\bibinfo {volume} {97}},\ \bibinfo {pages} {030501}
  (\bibinfo {year} {2015})}\BibitemShut {NoStop}%
\bibitem [{\citenamefont {Rad{\v z}i{\=u}t{\.e}}\ \emph
  {et~al.}(2014)\citenamefont {Rad{\v z}i{\=u}t{\.e}}, \citenamefont
  {Gaigalas}, \citenamefont {J{\"o}nsson},\ and\ \citenamefont
  {Biero{\'n}}}]{2014_Radziute_et_al_DHF_EDM}%
  \BibitemOpen
  \bibfield  {author} {\bibinfo {author} {\bibfnamefont {L.}~\bibnamefont
  {Rad{\v z}i{\=u}t{\.e}}}, \bibinfo {author} {\bibfnamefont {G.}~\bibnamefont
  {Gaigalas}}, \bibinfo {author} {\bibfnamefont {P.}~\bibnamefont
  {J{\"o}nsson}}, \ and\ \bibinfo {author} {\bibfnamefont {J.}~\bibnamefont
  {Biero{\'n}}},\ }\href@noop {} {\bibfield  {journal} {\bibinfo  {journal}
  {Phys. Rev. A}\ }\textbf {\bibinfo {volume} {90}} (\bibinfo {year}
  {2014})}\BibitemShut {NoStop}%
\bibitem [{\citenamefont {Latha}\ \emph {et~al.}(2009)\citenamefont {Latha},
  \citenamefont {Angom}, \citenamefont {Das},\ and\ \citenamefont
  {Mukherjee}}]{2009_Latha_et_al_Hg_EDM}%
  \BibitemOpen
  \bibfield  {author} {\bibinfo {author} {\bibfnamefont {K.~V.~P.}\
  \bibnamefont {Latha}}, \bibinfo {author} {\bibfnamefont {D.}~\bibnamefont
  {Angom}}, \bibinfo {author} {\bibfnamefont {B.~P.}\ \bibnamefont {Das}}, \
  and\ \bibinfo {author} {\bibfnamefont {D.}~\bibnamefont {Mukherjee}},\
  }\href@noop {} {\bibfield  {journal} {\bibinfo  {journal} {Phys. Rev. Lett.}\
  }\textbf {\bibinfo {volume} {103}},\ \bibinfo {pages} {083001} (\bibinfo
  {year} {2009})}\BibitemShut {NoStop}%
\bibitem [{\citenamefont {Latha}\ \emph {et~al.}(2015)\citenamefont {Latha},
  \citenamefont {Angom}, \citenamefont {Das},\ and\ \citenamefont
  {Mukherjee}}]{2015_Latha_et_al_Hg_EDM_Correction}%
  \BibitemOpen
  \bibfield  {author} {\bibinfo {author} {\bibfnamefont {K.~V.~P.}\
  \bibnamefont {Latha}}, \bibinfo {author} {\bibfnamefont {D.}~\bibnamefont
  {Angom}}, \bibinfo {author} {\bibfnamefont {B.~P.}\ \bibnamefont {Das}}, \
  and\ \bibinfo {author} {\bibfnamefont {D.}~\bibnamefont {Mukherjee}},\
  }\href@noop {} {\bibfield  {journal} {\bibinfo  {journal} {Phys. Rev. Lett.}\
  }\textbf {\bibinfo {volume} {115}},\ \bibinfo {pages} {059902(E)} (\bibinfo
  {year} {2015})}\BibitemShut {NoStop}%
\bibitem [{\citenamefont {Dzuba}\ \emph {et~al.}(2009)\citenamefont {Dzuba},
  \citenamefont {Flambaum},\ and\ \citenamefont
  {Porsev}}]{2009_Dzuba_Flambaum_Porsev_EDMvSM}%
  \BibitemOpen
  \bibfield  {author} {\bibinfo {author} {\bibfnamefont {V.~A.}\ \bibnamefont
  {Dzuba}}, \bibinfo {author} {\bibfnamefont {V.~V.}\ \bibnamefont {Flambaum}},
  \ and\ \bibinfo {author} {\bibfnamefont {S.~G.}\ \bibnamefont {Porsev}},\
  }\href@noop {} {\bibfield  {journal} {\bibinfo  {journal} {Phys. Rev. A}\
  }\textbf {\bibinfo {volume} {80}} (\bibinfo {year} {2009})}\BibitemShut
  {NoStop}%
\bibitem [{\citenamefont {Dzuba}\ \emph {et~al.}(2002)\citenamefont {Dzuba},
  \citenamefont {Flambaum}, \citenamefont {Ginges},\ and\ \citenamefont
  {Kozlov}}]{2002_Dzuba_et._al._EDMvsSM}%
  \BibitemOpen
  \bibfield  {author} {\bibinfo {author} {\bibfnamefont {V.~A.}\ \bibnamefont
  {Dzuba}}, \bibinfo {author} {\bibfnamefont {V.~V.}\ \bibnamefont {Flambaum}},
  \bibinfo {author} {\bibfnamefont {J.~S.~M.}\ \bibnamefont {Ginges}}, \ and\
  \bibinfo {author} {\bibfnamefont {M.~G.}\ \bibnamefont {Kozlov}},\
  }\href@noop {} {\bibfield  {journal} {\bibinfo  {journal} {Phys. Rev. A}\
  }\textbf {\bibinfo {volume} {66}},\ \bibinfo {pages} {012111} (\bibinfo
  {year} {2002})}\BibitemShut {NoStop}%
\bibitem [{\citenamefont {Dmitriev}\ and\ \citenamefont
  {Sen'kov}(2003)}]{2003_Dmitriev_and_Sen'kov_Hg_Schiff_Moment}%
  \BibitemOpen
  \bibfield  {author} {\bibinfo {author} {\bibfnamefont {V.~F.}\ \bibnamefont
  {Dmitriev}}\ and\ \bibinfo {author} {\bibfnamefont {R.~A.}\ \bibnamefont
  {Sen'kov}},\ }\href@noop {} {\bibfield  {journal} {\bibinfo  {journal} {Phys.
  Rev. Lett.}\ }\textbf {\bibinfo {volume} {91}},\ \bibinfo {pages} {212303}
  (\bibinfo {year} {2003})}\BibitemShut {NoStop}%
\bibitem [{\citenamefont {de~Vries}\ \emph {et~al.}(2015)\citenamefont
  {de~Vries}, \citenamefont {Mereghetti},\ and\ \citenamefont
  {Walker-Loud}}]{2015_deVries_Mereghetti_Walker_CP_violation_in_CPT}%
  \BibitemOpen
  \bibfield  {author} {\bibinfo {author} {\bibfnamefont {J.}~\bibnamefont
  {de~Vries}}, \bibinfo {author} {\bibfnamefont {E.}~\bibnamefont
  {Mereghetti}}, \ and\ \bibinfo {author} {\bibfnamefont {A.}~\bibnamefont
  {Walker-Loud}},\ }\href@noop {} {\bibfield  {journal} {\bibinfo  {journal}
  {Phys. Rev. C}\ }\textbf {\bibinfo {volume} {92}},\ \bibinfo {pages} {045201}
  (\bibinfo {year} {2015})}\BibitemShut {NoStop}%
\bibitem [{\citenamefont {Bsaisou}\ \emph {et~al.}(2015)\citenamefont
  {Bsaisou}, \citenamefont {de~Vries}, \citenamefont {Hanhart}, \citenamefont
  {Liebig}, \citenamefont {Mei{\ss}ner}, \citenamefont {Minossi}, \citenamefont
  {Nogga},\ and\ \citenamefont
  {Wirzba}}]{2015_Bsaisou_et_al_Nucleon_EDMs_in_chiral_EFT}%
  \BibitemOpen
  \bibfield  {author} {\bibinfo {author} {\bibfnamefont {J.}~\bibnamefont
  {Bsaisou}}, \bibinfo {author} {\bibfnamefont {J.}~\bibnamefont {de~Vries}},
  \bibinfo {author} {\bibfnamefont {C.}~\bibnamefont {Hanhart}}, \bibinfo
  {author} {\bibfnamefont {S.}~\bibnamefont {Liebig}}, \bibinfo {author}
  {\bibfnamefont {U.-G.}\ \bibnamefont {Mei{\ss}ner}}, \bibinfo {author}
  {\bibfnamefont {D.}~\bibnamefont {Minossi}}, \bibinfo {author} {\bibfnamefont
  {A.}~\bibnamefont {Nogga}}, \ and\ \bibinfo {author} {\bibfnamefont
  {A.}~\bibnamefont {Wirzba}},\ }\href@noop {} {\bibfield  {journal} {\bibinfo
  {journal} {J. High Energy Phys.}\ }\textbf {\bibinfo {volume} {03}} (\bibinfo
  {year} {2015})},\ \bibinfo {note} {104}\BibitemShut {NoStop}%
\bibitem [{\citenamefont {Guo}\ \emph {et~al.}(2015)\citenamefont {Guo},
  \citenamefont {Horsley}, \citenamefont {Mei{\ss}ner}, \citenamefont
  {Nakamura}, \citenamefont {Perlt}, \citenamefont {Rakow}, \citenamefont
  {Schierholz}, \citenamefont {Schiller},\ and\ \citenamefont
  {Zanotti}}]{2015_Guo_et_al_nEDM_from_lattice_QCD}%
  \BibitemOpen
  \bibfield  {author} {\bibinfo {author} {\bibfnamefont {F.-K.}\ \bibnamefont
  {Guo}}, \bibinfo {author} {\bibfnamefont {R.}~\bibnamefont {Horsley}},
  \bibinfo {author} {\bibfnamefont {U.-G.}\ \bibnamefont {Mei{\ss}ner}},
  \bibinfo {author} {\bibfnamefont {Y.}~\bibnamefont {Nakamura}}, \bibinfo
  {author} {\bibfnamefont {H.}~\bibnamefont {Perlt}}, \bibinfo {author}
  {\bibfnamefont {P.~E.~L.}\ \bibnamefont {Rakow}}, \bibinfo {author}
  {\bibfnamefont {G.}~\bibnamefont {Schierholz}}, \bibinfo {author}
  {\bibfnamefont {A.}~\bibnamefont {Schiller}}, \ and\ \bibinfo {author}
  {\bibfnamefont {J.~M.}\ \bibnamefont {Zanotti}},\ }\href@noop {} {\bibfield
  {journal} {\bibinfo  {journal} {Phys. Rev. Lett.}\ }\textbf {\bibinfo
  {volume} {115}},\ \bibinfo {pages} {062001} (\bibinfo {year}
  {2015})}\BibitemShut {NoStop}%
\bibitem [{\citenamefont {Pospelov}(2002)}]{2002_Pospelov_quark_chromo_EDM}%
  \BibitemOpen
  \bibfield  {author} {\bibinfo {author} {\bibfnamefont {M.}~\bibnamefont
  {Pospelov}},\ }\href@noop {} {\bibfield  {journal} {\bibinfo  {journal}
  {Phys. Lett. B}\ }\textbf {\bibinfo {volume} {530}},\ \bibinfo {pages} {123 }
  (\bibinfo {year} {2002})}\BibitemShut {NoStop}%
\bibitem [{\citenamefont {Latha}\ \emph {et~al.}(2008)\citenamefont {Latha},
  \citenamefont {Angom}, \citenamefont {Chaudhuri}, \citenamefont {Das},\ and\
  \citenamefont {Mukherjee}}]{2008_Latha_et_al_Core_Polrization_EDM}%
  \BibitemOpen
  \bibfield  {author} {\bibinfo {author} {\bibfnamefont {K.~V.~P.}\
  \bibnamefont {Latha}}, \bibinfo {author} {\bibfnamefont {D.}~\bibnamefont
  {Angom}}, \bibinfo {author} {\bibfnamefont {R.~K.}\ \bibnamefont
  {Chaudhuri}}, \bibinfo {author} {\bibfnamefont {B.~P.}\ \bibnamefont {Das}},
  \ and\ \bibinfo {author} {\bibfnamefont {D.}~\bibnamefont {Mukherjee}},\
  }\href@noop {} {\bibfield  {journal} {\bibinfo  {journal} {J. Phys. B}\
  }\textbf {\bibinfo {volume} {41}},\ \bibinfo {pages} {035005} (\bibinfo
  {year} {2008})}\BibitemShut {NoStop}%
\bibitem [{\citenamefont {Chupp}\ and\ \citenamefont
  {Ramsey-Musolf}(2015)}]{2015_Chupp_Xe_EDM_Motivation}%
  \BibitemOpen
  \bibfield  {author} {\bibinfo {author} {\bibfnamefont {T.}~\bibnamefont
  {Chupp}}\ and\ \bibinfo {author} {\bibfnamefont {M.}~\bibnamefont
  {Ramsey-Musolf}},\ }\href@noop {} {\bibfield  {journal} {\bibinfo  {journal}
  {Phys. Rev. C}\ }\textbf {\bibinfo {volume} {91}},\ \bibinfo {pages} {035502}
  (\bibinfo {year} {2015})}\BibitemShut {NoStop}%
\end{thebibliography}%
\end{document}